\documentclass[journal,twoside,12ptl]{IEEEtran}
\IEEEoverridecommandlockouts
\usepackage{fancyhdr}
\usepackage{lineno}
\usepackage[ruled, lined, linesnumbered, commentsnumbered]{algorithm2e}
\usepackage[cmex10]{amsmath}
\usepackage{array, setspace}
\usepackage{mdwmath}
\usepackage{mdwtab}
\usepackage[usenames,dvipsnames]{color}
\usepackage{cases}
\usepackage{multirow, lscape}
\usepackage[ampersand]{easylist}
\usepackage{graphicx}
\usepackage{color}
\usepackage{cite}
\usepackage{graphicx}
\usepackage{graphicx, subfigure,floatrow}
\usepackage[cmex10]{amsmath}
\usepackage{esint}
\usepackage{amssymb}
\usepackage{cases}
\usepackage{algorithmic}
\usepackage{array}
\usepackage{epstopdf}
\usepackage{amsthm}
\usepackage{bm}
\usepackage{caption}
\usepackage{setspace}
\usepackage{stfloats}
\usepackage{verbatim}
\usepackage{doi}

\usepackage{amsmath}
\usepackage{amssymb}
\usepackage{mathptmx}
\usepackage{bm}

\usepackage{algorithmic}
\usepackage{eucal}
\usepackage{underscore}

\usepackage{mathrsfs}

\def\BibTeX{{\rm B\kern-.05em{\sc i\kern-.025em b}\kern-.08em
    T\kern-.1667em\lower.7ex\hbox{E}\kern-.125emX}}
\begin{document}
\title{Complementary Waveforms for Range-Doppler Sidelobe Suppression Based on a Null Space Approach\\
}
\author{\IEEEauthorblockN{ Jiahuan Wang, Pingzhi Fan, Des McLernon and Zhiguo Ding}
\thanks{This work was supported by NSFC Project No.62020106001/No.61731017, National Key R\&D Project No.2018YFB1801104, and 111 project No.111-2-14.}
\thanks{J. Wang and P. Fan are with the School of Information Science and Technology, Southwest Jiaotong University, Chengdu, China (e-mail: jiahuanwang@my.swjtu.edu.cn, pzfan@swjtu.edu.cn)}
\thanks{D. McLernon is with the School of Electronic and Electrical Engineering, University of Leeds (e-mail: D.C.McLernon@leeds.ac.uk)}
\thanks{Z. Ding is with the School of Electrical and Electronic Engineering, The University of Manchester (e-mail: zhiguo.ding@manchester.ac.uk)}
}

\maketitle
\begin{abstract}
While Doppler resilient complementary waveforms have previously been considered to suppress range sidelobes within a Doppler interval of interest in radar systems, their capability of Doppler resilience has not been fully utilized. In this paper, a new construction of Doppler resilient complementary waveforms based on a null space  is proposed. With this new construction, one can flexibly include a specified Doppler interval of interest or even an overall Doppler interval into a term which results in range sidelobes. We can force this term to zero, which can be solved to obtain a null space. From the null space, the characteristic vector to control the transmission of basic Golay waveforms,  and the coefficients of  the receiver filter for Golay complementary waveform can be extracted. Besides, based on the derived null space, two challenging non-convex optimization problems are formulated and solved for maximizing the signal-to-noise ratio (SNR).  Moreover, the coefficients of the receiver filter and  the characteristic vector can be applied to fully polarimetric radar systems to achieve nearly perfect Doppler resilient performance, and hence fully suppress the inter-antenna interferences.
\end{abstract}


\section{Introduction}
In pulsed radar systems, pulse compression technology\cite{farnett1990pulse}\cite{mahafza2002radar} has been commonly  used to obtain high pulse energy, large bandwidth, and improved range resolution. Through the use of a matched filter receiver, the returned signal reflected by a target goes through a filter matched to the reverse and conjugate version of the transmitted pulse. Then the echo signal is compressed into a short pulse which is shown in the matched filter output along with the well-known maximum SNR. However, the matched filter output also has undesired range sidelobes if the pulses are not carefully chosen. The range sidelobes of a strong target may mask main the peak of a weak target near the strong target. Therefore, low range sidelobes are desirable when dealing with multi-targets. 

In order to obtain these low range sidelobes, phase coding is usually used in radar for digital pulse compression. For a phase coded waveform, it is phase coded by a unimodular code or sequence. The matched filter output of a phase coded waveform is controlled by an aperiodic auto-correlation function of a code (sequence).
For bi-phase codes, the Barker code is a famous code whose aperiodic auto-correlation function has low sidelobes with only one element amplitude value. In addition, polyphase codes proposed by  Heimiller\cite{heimiller1961phase} and Chu\cite{chu1972polyphase} also have low sidelobes of aperiodic auto-correlation functions. However, it is impossible to achieve zero sidelobes of an aperiodic auto-correlation with one unimodular sequence\cite{pezeshki2008doppler}. This has resulted in the use of Golay complementary pairs in phase coding.

 In radar, Golay waveforms phase coded by Golay pairs are coherently transmitted, and the returned signals are also coherently processed through the matched filter. Then the sum of the matched filter outputs has no range sidelobes since Golay pairs have an impulse-like aperiodic autocorrelation function. But the Golay pairs are sensitive to Doppler effects which result from the moving targets. In other words, as the  inter-pulse Doppler shift changes the phase of the complementary waveforms, the matched filter outputs' sum of complementary  waveforms have fairly high range sidelobes.  In order to solve this problem, some methods for constructing Doppler resilient waveforms have been proposed. These existing construction methods fall into two categories.
 
 The first category only concerns the transmission of the basic Golay waveforms. The transmission is determined by space-time codes. Examples of these codes that play a key role in constructing Doppler resilient Golay waveforms (pulse train) are first-order Reed-M{\"u}ller codes\cite{suvorova2007doppler}, Prouhet–Thue–Morse (PTM) sequences\cite{pezeshki2008doppler}, oversampled PTM sequences\cite{chi2009range},  generalized PTM sequences\cite{tang2014construction}, and equal sums of powers (ESP) sequences\cite{nguyen2016doppler}. The first-order Reed-M{\"u}ller codes decrease the range sidelobes (less than -60dB) at a specific Doppler value (e.g., 0.25 rad). PTM sequences can almost clear range sidelobes (which are approximately equal to -80dB) near zero Doppler (i.e., $[-0.1,0.1]$ in rad). The oversampled PTM sequences can also almost clear range sidelobes not only near zero Doppler but also in all rational Doppler shifts (in rad). The generalized PTM sequence is generally used for a complementary set, but it is also compatible with the traditional PTM sequence when only a complementary pair is considered. Therefore, the generalized PTM sequence for a complementary waveform set can almost clear range sidelobes near zero Doppler. The ESP sequence has almost the same Doppler resilient performance compared to a PTM sequence. In fact, ESP and PTM are closely related to the solutions of the Prouhet-Tarry-Escott (PTE) problem. However, ESP sequences need two antennas to transmit the Golay waveforms in some pulse repetition intervals (PRIs) that will result in inter-waveform interferences. 
 
 The second category focuses on not only the transmission  of the basic Golay waveforms but also the coefficients of the receiver filter.  The pulse weighing technology is similar to traditional time-domain window design. In the windowing method, the coefficients of the matched filters are determined from known time-domain windows such as the rectangular window, B-spline windows, the triangular window, Hann and Hamming windows, which can achieve range sidelobe suppression. However, when Doppler effects are present, these well-known traditional windows cannot be directly used to suppress the sidelobes down to a very low level. Wu $et$ $al.$\cite{wu2020range} jointly considered the coefficients of the receiver and a given window function (e.g., Hamming and rectangular windows) with the higher-order Doppler null and  max-SNR constraints, so that the traditional window function can be indirectly used to suppress the range sidelobes in a given Doppler interval. Generally, these known windows are not suitable for Doppler resilience. Dang $et\, al$. \cite{dang2011coordinating}\cite{dang2014signal}\cite{dang2020coordinating} proposed the binomial design (BD) that puts binomial coefficients as the coefficients of the receiver filter, and alternatively transmits Golay waveforms at the transmitter. The BD method has a relatively large Doppler resilient interval, in which range sidelobes are suppressed down to almost zero.

So in this paper, the construction of Doppler resilient complementary waveforms based on a null space approach is proposed. 
The main contributions of this paper are listed as follows:
\begin{itemize}
 \item For the defined ambiguity function \cite{dang2020coordinating}, it is proved theoretically that there exists a totally Doppler resilient type ambiguity function, as well as  the delay resilient type ambiguity function,  under certain conditions by solving a linear system and finding the null space. 
\item A Doppler resilient transmission waveform based on Golay pairs which can suppress the range sidelobes in  a specified Doppler interval of interest, or even overall Doppler interval, is designed. Here, the design problem is formulated as a linear system with one key term which results in range sidelobes. 
\item By forcing the above key term in the formulated linear system to zero, the linear system can be solved to obtain a null space. From the null space, the characteristic vector to control the transmission of basic Golay waveforms, and the coefficients of the receiver filter, in the intervals of interest or the entire Doppler interval, are extracted. 
\item Based on the derived null space, a complex optimization problem, which is non-convex and non-concave in nature, is formulated for maximizing the signal-to-noise ratio (SNR). A heuristic coordinated descent (HCD) algorithm is proposed to obtain a sub-optimal solution of the formulated optimization problem, where the null space is taken as a means to transform the problem into  a simpler one, i.e, finding the coefficients of a linear combination of null space basis vectors.
\item The above characteristic vector in the transmission waveform, and the coefficients of the receiver filter, when applied to  fully polarimetric radar systems, can achieve nearly perfect Doppler resilient performance and fully suppress the inter-antenna interference.
\item The delay resilient problems are also investigated and solved by using the frequency-domain phase coding, also based on the null space algorithm. In fact, the null space algorithm is used to obtain the frequency-domain characteristic vector, and also the coefficients of the frequency-domain filter at the receiver. \end{itemize}

\subsection{Notation}
The superscripts $(\cdot)^{T},(\cdot)^{*}$ and $(\cdot)^{H}$ denote transpose, complex conjugate, and conjugate transpose, respectively. In addition, `$\circ$' denotes the Hadamard product and $\mathrm{Null}(\mathbf{E})$ denotes null space of matrix $\mathbf{E}$.

 \section{Signal Model}
\subsection{Time-Domain Signal Model}
A pair of biphase sequences $\mathbf{x}$ and $\mathbf{y}$ is called Golay pair or complementary pair if
\begin{eqnarray}
	C_\mathbf{x}[k]+C_\mathbf{y}[k]=2L\delta_k, \,\, k=-L+1,\cdots,0,\cdots, L-1
\end{eqnarray}
where  $C_\mathbf{x}[k]$ is the auto-correlation function of $\mathbf x$ at lag $k$, $\delta_k$ is the Kronecker delta function, and $\mathbf{x}=[x[0],x[1],\cdots,x[L-1]]^T$, $\mathbf{y}=[y[0],y[1],\cdots,y[L-1]]^T$.

In the signal model, the basic Golay complementary waveforms $s_x(t)$ and $s_y(t)$ are phase coded by the Golay complementary pair $(\mathbf{x},\mathbf{y})$\cite{chi2009range,dang2011coordinating,dang2014signal}, i.e.,
\begin{eqnarray}
	s_x(t)=\sum_{l=0}^{L-1}x[l]u(t-lT_c),\,\, s_y(t)=\sum_{l=0}^{L-1}y[l]u(t-lT_c),
\end{eqnarray}
where $u(t)$ is a unit-energy baseband pulse shape, and $T_c$ is the chip length.

Let the vector $\mathbf{p}=[p_0,p_1,\cdots,p_{N-1}]^T$ be the characteristic vector to control the transmission of basic Golay waveforms $s_x(t)$ and $s_y(t)$, where $N$ is the number of pulses and $p_n=1$ or $-1$.  If $p_n=1$,  $s_x(t)$ is transmitted, otherwise $s_y(t)$ is transmitted. Thus the characteristic vector $\mathbf{p}$ and the basic Golay waveforms,  $s_x(t)$ and $s_y(t)$ constitute the Golay transmission waveform or complementary waveform, $Z_p(t)$, i.e.,
\begin{eqnarray}
	Z_{P}(t)=\frac{1}{2} \sum_{n=0}^{N-1}\left[\left(1+p_{n}\right) s_{x}(t-n T)+\left(1-p_{n}\right) s_{y}(t-n T)\right],\label{eq:Zp}
\end{eqnarray}
where $T$ denotes the pulse repetition interval (PRI). 

Let the input to the matched filter be $Z_P (t) e^{j\nu t}$, where $\nu=2\pi f_d$, and $f_d$ is the Doppler shift in Hz. Also, $Z_P (t) e^{j\nu t}$ passes through the linear filter  with impulse response $Z_{W}^*(-t)$, where
\begin{eqnarray}
\begin{split}
	Z_{W}(t)=\frac{1}{2} \sum_{n=0}^{N-1}w_n^*&\left[\left(1+p_{n}\right) s_{x}(t-n T)\right.\\
	&\left.+\left(1-p_{n}\right) s_{y}(t-n T)\right],\label{eq:Zw}
	\end{split}
\end{eqnarray}
$w_n\in \mathbb{C}$ is the coefficient of receiver filter and $\mathbf{w} = [w_0,w_1,\cdots,w_{N-1}]^T$.

Then the output, i.e., the cross-ambiguity function is given by
\begin{eqnarray}
	\chi_{P, W}(\tau, \nu)=\int_{-\infty}^{+\infty} Z_{P}(t) Z_{W}^{*}(t-\tau) \mathrm{e}^{j \nu t} d t.\label{eq:chiPW}
\end{eqnarray}

The radar parameters (such as chip length $T_c$ and PRI $T$) are chosen to ensure that  $L T_c$ is much less than $T$ and $\nu T$ is almost equal to zero.
After carefully choosing the radar parameters, the center lobe of $\chi_{P, W}(\tau, \nu)$ depends on the discrete cross ambiguity function\cite{wu2020range}\cite{dang2020coordinating}
\begin{equation}
	\begin{aligned} \mathcal{A}_{P, W}(k, \theta) &=\frac{1}{2}\left[C_{x}[k]+C_{y}[k]\right] \sum_{n=0}^{N-1} w_n e^{j n \theta} \\ &+\frac{1}{2}\left[C_{x}[k]-C_y[k]\right] \sum_{n=0}^{N-1} p_{n} w_n e^{j n \theta}, \end{aligned}\label{eq:APW}
\end{equation}
where $\theta = \nu T = 2\pi f_d T$ is the Doppler shift in radians.

The first part of (\ref{eq:APW}) only determines the shape of $\mathcal{A}_{P, W}(0, \theta)$ since the Golay complementary pair ($\mathbf{x}$,$\mathbf{y})$ makes the first formula of (\ref{eq:APW}) vanish at nonzero $k$. However, the second part of (\ref{eq:APW}) determines the range sidelobes around the Doppler shift $\theta$, in which the choices of $p_n$ and $w_n$ are important. In \cite{dang2020coordinating}\cite{wu2020range}, two performance metrics (i.e., Doppler resilience and SNR) are chosen to judge the Doppler resilient complementary waveform specified by $\{\mathbf{p}, \mathbf{w}\}$.

\subsection{Frequency-Domain Pulse Amplitude Modulation (PAM) Signal Model}
The frequency-domain PAM waveforms for $\mathbf{x}$ and $\mathbf{y}$ are given by \cite{pezeshki2009sidelobe}\cite{wang2019range}
\begin{eqnarray}
	\hat{x}(\omega) = \sum_{l=0}^{L-1}x[l]\Omega(\omega-lW_c),\,\,
	\hat{y}(\omega) = \sum_{l=0}^{L-1}y[l]\Omega(\omega-lW_c),
\end{eqnarray}
where $W_c$ denotes the subcarrier spacing and $\Omega(\omega)$ denotes the subcarrier complex amplitude.

Then the transmitted frequency PAM pulse train is expressed as
\begin{eqnarray}
\hat{z}_{\mathcal{P}} (\omega) = \sum_{n=0}^{N-1}p_n\hat{x}(\omega-nW_0)+(1-p_n)\hat{y}(\omega-nW_0),
\end{eqnarray}
and the impluse response of the filter in the frequency domain is $\hat{z}_{\mathcal{Q}} (-\omega)$, where
\begin{eqnarray}
\hat{z}_{\mathcal{Q}} (\omega) = \sum_{n=0}^{N-1}w_n^*[p_n\hat{x}(\omega-nW_0)+(1-p_n)\hat{y}(\omega-nW_0)],
\end{eqnarray}	
where $W_0$ is the frequency-domain PRI, where $W_0\gg W_c$.

Then the cross ambiguity function of $\hat{z}_\mathcal{P}(\omega)$ and $\hat{z}_\mathcal{Q}(\omega)$ is given by
\begin{eqnarray}
\chi_{P,W}^f (\nu,\tau) = \int_{-\infty}^{+\infty} \hat{z}_\mathcal{P}(\omega) \hat{z}_\mathcal{Q}^*(\omega-\nu)e^{-j\tau \omega}d\omega.
\end{eqnarray}
After applying the inverse  continuous-time Fourier transform, the transmitted frequency domain PAM waveform $\hat{z}_{\mathcal{P}} (\omega)$ is an OFDM waveform in the time-domain:

\begin{eqnarray}
\begin{split}
Z_\mathcal{P} (t) &= \frac{1}{2 \pi} \sum_{n=0}^{N-1}  \sum_{l=0}^{L-1}( p_n {x}[l]+(1-p_n){y}[l]) \\
& \quad \cdot e^{j(nW_0+lW_c)t}\hat{\Omega}(t), \label{eq: ZpOFDM}
\end{split}
\end{eqnarray}
and similarly the inverse Fourier transform of $\hat{z}_{\mathcal{Q}} (\omega)$ is given by
\begin{eqnarray}\label{eq: ZqODFM}
\begin{split}
Z_\mathcal{Q} (t) &= \frac{1}{2 \pi} \sum_{n=0}^{N-1}  \sum_{l=0}^{L-1}w_n( p_n {x}[l]\\
&{}\quad+(1-p_n){y}[l]) e^{j(nW_0+lW_c)t}\hat{\Omega}(t),
\end{split}
\end{eqnarray}
where $\hat{\Omega}(t) = \mathcal{F}^{-1} \left\{2\pi \Omega(\omega) \right\}$ and $\mathcal{F}^{-1} $ denotes the inverse Fourier transform.

The discrete ambiguity function based on $\chi_{P,W}^f(\nu,\tau)$ is given by
\begin{equation}
\begin{aligned}
\mathcal{B}_{P,W}(i,\alpha)=& \frac{1}{2}\left[C_{\hat{x}}[i]+C_{\hat{y}}[i]\right] \sum_{n=0}^{N-1} w_n e^{j n \alpha} \\
&+\frac{1}{2}\left[C_{\hat{x}}[i]-C_{\hat{y}}[i]\right] \sum_{n=0}^{N-1}(-1)^{p_n} w_n e^{j n \alpha},
\end{aligned}\label{eq:BPW}
\end{equation}
where $\alpha = \tau W_0$ is the time shift.

\section{Doppler/delay resilience based windowing}
\subsection{Doppler Resilience}

\emph{Definition 1:}
The ambiguity function $\mathcal{A}_{P, W}(k, \theta)$ is a Doppler resilient type if the following conditions hold:
\begin{eqnarray}
	 |\mathcal{A}_{P, W}(0,\theta)|> 0, \quad \theta\in \Theta,\label{eq:purpose}
\end{eqnarray}	
and
\begin{eqnarray}
	 |\mathcal{A}_{P, W}(k, \theta)/\mathcal{A}_{P, W}(0, \theta)|\leq \eta, \quad k\neq 0, \,\,\theta\in \Theta,\label{eq:D2}
\end{eqnarray}	
where $\Theta=[0,D_I]$ is the specified  Doppler interval of interest, $D_I$ is a positive real number, and $\eta$ is a very small positive real number.

We will show that the range sidelobes may almost vanish around the specified   Doppler interval of interest $\Theta = [0,D_I]$ , and maintain $|\mathcal{A}(0,\theta)|\neq 0$. The best case is that $\eta$ in (\ref{eq:D2}) is totally equal to zero such that the discrete ambiguity function satisfies the following equation
\begin{eqnarray}
	\mathcal{A}_{P, W}(k, \theta)=\mathcal{A}_{P, W}(0,\theta)\delta_k, \quad \theta\in \Theta\label{eq:purpose}.
\end{eqnarray}

\emph{Definition 2:}
The ambiguity function is a totally Doppler resilient type for $\theta\in \Theta$ if 
\begin{eqnarray}
	\mathcal{A}_{P, W}(k, \theta)=\mathcal{A}_{P, W}(0,\theta)\delta_k,\,\, \theta\in \Theta,
\end{eqnarray}
where $|\mathcal{A}_{P, W}(0,\theta)|> 0$.

To make analyzing the ambiguity function much easier, a Doppler Vandermonde matrix $\mathbf{E}$ is proposed as follows:
\begin{eqnarray}
\mathbf{E} =
\left[\begin{matrix}
  e^{j0\theta_0}& e^{j1\theta_0}&e^{j2\theta_0}&\cdots&e^{j(N-1)\theta_0} \\
  e^{j0\theta_1}& e^{j1\theta_1}&e^{j2\theta_1}&\cdots&e^{j(N-1)\theta_1} \\
  \vdots&\vdots&\vdots&\vdots&\vdots\\
  e^{j0\theta_{M-1}}& e^{j1\theta_{M-1}}&e^{j2\theta_{M-1}}&\cdots&e^{j(N-1)\theta_{M-1}} \\
\end{matrix}\right].\label{eq:E}
\end{eqnarray}
where $\Theta_{\Delta} = \{\theta_0,\theta_1,\cdots,\theta_{M-1}\}$ and $\Theta_{\Delta}\subset \Theta$.

\emph{Proposition 1:}
	If $(\mathbf{x},\mathbf{y})$ is a Golay pair, then the discrete ambiguity function defined in (\ref{eq:APW}) can be rewritten as
	\begin{eqnarray}
		\mathcal{A}_{P,W} (k,\theta)= 	\left\{
		\begin{aligned}
			&\frac{1}{2}(C_{\mathbf{x}}(k)-C_{\mathbf{y}}(k))\sum_{n=0}^{N-1}p_n w_n e^{jn\theta}, &\,\, k\neq0,\\
			&L\sum_{n=0}^{N-1}w_n e^{jn\theta},  &\,\, k= 0.
			\end{aligned}
		\right.
	\end{eqnarray}

\begin{proof}
 When $k\neq 0$, since $(\mathbf{x},\mathbf{y})$ is a Golay pair, then $C_{\mathbf{x}}[k]+C_{\mathbf{y}}[k]=0$, so that 
 \begin{equation}
 	\mathcal{A}_{P,W} (k,\theta)=\frac{1}{2}(C_{\mathbf{x}}(k)-C_{\mathbf{y}}(k))\sum_{n=0}^{N-1}p_n w_n e^{jn\theta}.
 \end{equation}
 When $k= 0$, $C_{\mathbf{x}}(0)=C_{\mathbf{y}}(0)=L$, and then it holds that
  \begin{equation}
 	\mathcal{A}_{P,W} (k,\theta)=L\sum_{n=0}^{N-1}w_n e^{jn\theta}.
 \end{equation}	
 \end{proof}

\emph{Remark 1:}
$\frac{1}{2}(C_{\mathbf{x}}(k)-C_{\mathbf{y}}(k))\sum_{n=0}^{N-1}p_n w_n e^{jn\theta}$ denotes the range sidelobes,  which we should clear.

\emph{Lemma 1:}
	Let $z_n= p_n w_n$, $n=0,1,2,\cdots,N-1$.  Then $|\mathcal{A}_{P,W}(k,\theta)|$ is bounded by $| f_{\mathbf{z}}(\theta)|$,  i.e.,
	\begin{equation}
		\begin{aligned} 
			|\mathcal{A}_{P,W}(k,\theta)| \leq L |f_{\mathbf{z}}(\theta)|, \quad k\neq 0,
		\end{aligned}
	\end{equation}
	where the key term $f_{\mathbf{z}}(\theta)$ is given by
	\begin{eqnarray}
		f_{\mathbf{z}}(\theta)= 	\sum_{n=0}^{N-1} z_n e^{j n \theta}.
	\end{eqnarray}
\begin{proof}
	Since $|C_{\mathbf{x}}(k)|\leq L$ and $|C_{\mathbf{y}}(k)|\leq L$ , then 
\begin{eqnarray}
		\frac{1}{2}|C_{\mathbf{x}}(k)-C_{\mathbf{y}}(k)|\leq L,\,\, \mathrm{for\,\, all}\,\,  k\in [-(L-1),L-1].
\end{eqnarray}
When $k\neq 0$, we have
		\begin{equation}
		\begin{aligned} 
		|\mathcal{A}_{P,W}(k,\theta)| 	&=|\frac{1}{2}\left[C_{x}(k)-C_{y}(k)\right] \sum_{n=0}^{N-1}z_n e^{j n \theta}| \\
			&\leq 	|\frac{1}{2}\left[C_{x}(k)-C_{y}(k)\right] ||\sum_{n=0}^{N-1}z_n e^{j n \theta}| \\
			&=L|\sum_{n=0}^{N-1}z_n e^{j n \theta}|\\
			&=L|f_{\mathbf{z}}(\theta)|.
		\end{aligned}
	\end{equation}
\end{proof}
	From lemma 1,  we konw that range sidelobes can be controlled by $f_{\mathbf{z}}(\theta)$. Therefore, clearing the range sidelobes means  $f_{\mathbf{z}}(\theta)=0$. 

\emph{Lemma 2:}
	$	f_{\mathbf{z}}(\theta)=0$, $\theta\in\Theta_{\Delta}$
	if and only if 
	$$ \mathbf{z}=[z_0,z_1,z_2,\cdots,z_{N-1}]^T\in \mathrm{Null}(\mathbf{E}),$$ where $z_n = p_n w_n$.

\begin{proof}
$	f_{\mathbf{z}}(\theta)=0$, $\theta\in\Theta_{\Delta}$ iff
	\begin{eqnarray}
		\sum_{n=0}^{N-1} z_n e^{j n \theta}=0,  \quad \mathrm{for\,\, all\,}\theta \in \Theta.\label{eq:obj1}
	\end{eqnarray}
i.e.,
	\begin{eqnarray}
\left[\begin{matrix}
   \sum_{n=0}^{N-1} z_n e^{j n \theta_0}\\
   \sum_{n=0}^{N-1} z_n e^{j n \theta_1}\\
   \vdots\\
   \sum_{n=0}^{N-1} z_n e^{j n \theta_{M-1}}
\end{matrix}\right]	=\bf 0,
\end{eqnarray}

\begin{eqnarray}
\implies	\mathbf{E}\mathbf{z}=\bf 0.
\end{eqnarray}
	In other words, $\mathbf{z} \in \mathrm{Null}(\mathbf{E})$.
\end{proof}

\emph{Theorem 1:}
	If $\mathbf{p}$ and $\mathbf{w}$ satisfy $\mathbf{w} \neq \mathbf{0}$, $\mathbf{w} \notin \mathrm{Null}(\mathbf{E})$ and $\mathbf{p}\circ \mathbf{w} \in \mathrm{Null}(\mathbf{E})$,
	 the ambiguity function $\mathcal{A}_{P,W}(k,\theta)$ is a totally Doppler resilient type ambiguity function, i.e.,
	\begin{eqnarray}
		\mathcal{A}_{P,W}(k,\theta) = 0, \quad \quad  k\neq 0, \,\, \theta \in \Theta_{\Delta}.
	\end{eqnarray}

\begin{proof}
	This is easily obtained from Lemma 1 and Lemma 2.
\end{proof}

Based on Theorem 1, the range sidelobes in $\Theta_{\Delta}$ will vanish. However, the  Doppler interval of interest $\Theta=[0,D_I]$ or the overall Doppler interval $[0,\pi]$ is truely focused, within which the range sidelobes are suppressed. In other words, we hope the range sidelobes of the ambiguity function $\mathcal{A}_{P,W}(k,\theta)$ can be no more than -90dB in the Doppler interval of interest $[0,D_I]$. 

\emph{Proposition 2:}
	In the Doppler interval of interest $[0,D_I]$, the discrete Doppler shifts can be chosen as $\theta_m = m  D_I/(M-1)$, $m=0,1,\cdots,M-1$. If $f_\mathbf{z}(\theta_m)=0$, $\theta_m=0,1,\cdots,M-1$, then the range sidelobes of $\mathcal{A}_{P,W}(k,\theta)$ can be suppressed for all $\theta \in [0,D_I]$, i.e., $\mathcal{A}_{P,W}(k,\theta)\rightarrow 0$.

After solving the linear system $\mathbf{Ez}=\bf 0$, we find the null space of $\mathbf{E}$. Then $\mathbf{p}$ and $\mathbf{w}$ can be found based on $\text{Null}(\mathbf{E})$. Supposed that $\hat{\mathbf{z}} \in \text{Null}(\mathbf{E})$, then $\mathbf{p}$ and $\mathbf{w}$ are solved as follows:
\begin{equation}
p_n=
\left\{
             \begin{array}{lr}
             +1, & \text{if} \quad \mathrm{Re}\{\hat{\mathbf{z}}\}\geq 0, \\
             -1, & \text{if} \quad \mathrm{Re}\{\hat{\mathbf{z}}\}< 0,\\
             \end{array}
\right.\label{eq:p}
\end{equation}

\begin{equation}
w_n=
\left\{
             \begin{array}{lr}
             +\hat{z}_n, & \text{if} \quad \mathrm{Re}\{\hat{\mathbf{z}}\}\geq 0, \\
             -\hat{z}_n, & \text{if} \quad \mathrm{Re}\{\hat{\mathbf{z}}\}< 0.\\
             \end{array}
\right.\label{eq:w}
\end{equation}		
It is easy to verify that $\mathbf{w} \neq \mathbf{0}$, $\mathbf{w} \notin \mathrm{Null}(\mathbf{E})$. Then we design an algorithm for finding $\mathbf{p}$ and $\mathbf{w}$ in algorithm 1.
\begin{algorithm}[htb]
\setstretch{1.5}
\caption{Null space (NS) algorithm for obtaining $\mathbf{p}$ and $\mathbf{w}$:}
\label{alg:Framwork}
\begin{algorithmic}[1] %
\STATE Input $N$, $D_I$, and $\theta_m= m  D_I/(M-1)$, \\
$m=0,1,\cdots, M-1$. Here, $M = N-1$.
\STATE Generate matrix $\mathbf{E}$ shown in (\ref{eq:E}).
\STATE Compute the null space of $\mathbf{E}$, i.e., $\text{Null}(\mathbf{E})$.
\STATE Select a solution from $\text{Null}(\mathbf{E})$, called $\hat{\mathbf{z}}$.
\STATE Obtain $\mathbf{p}$ and $\mathbf{w}$ as (\ref{eq:p}) and (\ref{eq:w})
\end{algorithmic}
\end{algorithm}

\emph{Remark 2:}
	In the practical operations of Algorithm 1, we use the MATLAB instruction ``null(E)'' to obtain the basis vectors (which can span the null space). Then we select a solution from $\mathrm{Null}(\mathbf{E})$. Usually, we can select the basis vector as our solution. If SNR is considered, we should carefully choose the vector by some algorithms which is also introduced in this paper.

\emph{Remark 3:}
	The number $M$ of the discrete Doppler shifts $\theta_m$ $(m=0,1,\cdots,M-1)$ is limited by the number of pulses $N$. This fact results from the solutions of the linear equations, i.e,  if $M<N$, $\mathbf{Ez}=\mathbf{0}$ must have nontrivial solutions. Therefore, $M$ can be chosen as $M = N-1$. Of course, the smaller the Doppler interval and the more the number of Doppler shifts,  the better suppression of range sidelobes  of $\mathcal{A}_{P,W}(k,\theta)$.

\subsection{Delay Resilience}
\emph{Definition 3:}
The ambiguity function $\mathcal{B}_{P, W}(i,\alpha)$ is a delay resilient type if the following formulas hold:
\begin{eqnarray}
	 |\mathcal{B}_{P, W}(0,\alpha)|> 0, \quad \alpha \in \Gamma,\label{eq:purpose}
\end{eqnarray}	
and
\begin{eqnarray}
	 |\mathcal{B}_{P, W}(i, \alpha)/\mathcal{B}_{P, W}(i, 0)|\leq \eta, \quad i\neq 0, \,\,\alpha\in \Gamma,\label{eq:purpose}
\end{eqnarray}	
where $\Gamma=[0,T_I]$ is the delay interval of interest, $T_I$ is a positive real number, and $\eta$ is a very small positive real number.

\emph{Definition 4:}
The ambiguity function $\mathcal{B}_{P, W}(i,\alpha)$ is a totally delay resilient type for $\theta\in \Theta$ if 
\begin{eqnarray}
	\mathcal{B}_{P, W}(i, \alpha)=\mathcal{B}_{P, W}(i,0)\delta_i,
\end{eqnarray}
where $|\mathcal{B}_{P, W}(i,0)|> 0$.

In order to simplify the analysis of the ambiguity function, a  delay Vandermonde matrix $\mathbf{T}$ is proposed as follows:
\begin{eqnarray}
\mathbf{T} =
\left[\begin{matrix}
  e^{j0\alpha_0}& e^{j1\alpha_0}&e^{j2\alpha_0}&\cdots&e^{j(N-1)\alpha_0} \\
  e^{j0\alpha_1}& e^{j1\alpha_1}&e^{j2\alpha_1}&\cdots&e^{j(N-1)\alpha_1} \\
  \vdots&\vdots&\vdots&\vdots&\vdots\\
  e^{j0\alpha_{M-1}}& e^{j1\alpha_{M-1}}&e^{j2\alpha_{M-1}}&\cdots&e^{j(N-1)\alpha_{M-1}} \\
\end{matrix}\right].\label{eq:T}
\end{eqnarray}
where $\Gamma_{\Delta}  = \{\alpha_0,\alpha_1,\cdots,\alpha_{M-1}\}$, and $\Gamma_{\Delta}\subset \Gamma=[0,T_I]$.

\emph{Lemma 3:}
	The first term of $\mathcal{B}_{P,W}(i,\alpha)$ in (\ref{eq:BPW}) is a delta function for any given $\theta \in \Theta$, i.e.,
	\begin{equation}
	\begin{aligned} \frac{1}{2}\left[C_{x}[i]+C_{y}[i]\right] \sum_{n=0}^{N-1} w_n e^{j n \alpha} = L(\sum_{n=0}^{N-1} w_n e^{j n \alpha})\delta_i,
	\end{aligned}
	\end{equation}
if and only if  $\mathbf{w} \neq \mathbf{0}$ and $\mathbf{w} \notin \mathrm{Null}(\mathbf{T})$.

\emph{Lemma 4:}
	The second term of $\mathcal{B}_{P,W}(i,\alpha)$ in (\ref{eq:BPW}) is zero for all given $\alpha \in \Gamma_{\Delta}$, i.e.,
	\begin{equation}
	\begin{aligned} \frac{1}{2}\left[C_{x}(i)-C_{y}(i)\right] \sum_{n=0}^{N-1}p_n w_n e^{j n \alpha} = 0,
	\end{aligned}
	\end{equation}
	if and only if $\mathbf{p}\circ \mathbf{w} \in \mathrm{Null}(\mathbf{T})$.
\begin{proof}
Since $C_{\mathbf{x}}(i)-C_{\mathbf{y}}(i)\neq 0$, when $k\neq 0$,then the following equation should hold:
	\begin{eqnarray}
		\sum_{n=0}^{N-1} p_{n} w_n e^{j n \alpha}=0,  \quad \mathrm{for\,\, all\,}\alpha \in \Gamma_{\Delta},
	\end{eqnarray}
i.e.,
\begin{eqnarray}
	\mathbf{T}\mathbf{z}=\bf 0,
\end{eqnarray}
where $\mathbf{z} = [z_0,z_1,\cdots,z_{N-1}]^T$, $z_n = p_n w_n$, $n=0,1,\cdots,N-1$,
$\theta_m \in \Theta$, $m=0,1,\cdots,M-1$.
	In other words, $\mathbf{p}\circ \mathbf{w} \in \mathrm{Null}(\mathbf{E})$.
\end{proof}

\emph{Theorem 2:}
	If $\mathbf{p}$ and $\mathbf{w}$ satisfy $\mathbf{w} \neq \mathbf{0}$, $\mathbf{w} \notin \mathrm{Null}(\mathbf{T})$ and $\mathbf{p}\circ \mathbf{w} \in \mathrm{Null}(\mathbf{T})$,
	then the ambiguity function $\mathcal{A}_{P,W}(k,\theta)$ is a totally delay resilient type ambiguity function, i.e., 
		\begin{eqnarray}
		\mathcal{B}_{P,W}(i,\alpha) = 0, \quad \quad  k\neq 0, \,\, \theta \in \Gamma_{\Delta}.
	\end{eqnarray}

\subsection{Signal-to-Noise Ratio (SNR)}
The SNR\cite{dang2011coordinating,dang2014signal,dang2020coordinating} is described as
\begin{eqnarray}
	\mathrm{SNR} = \frac{L\sigma_b^2}{N_0}\frac{\|\mathbf{w}\|_1^2}{\|\mathbf{w}\|_2^2},\label{eq:SNR}
\end{eqnarray}
where $\sigma_b^2$ is the power of the target and $N_0$ is the power spectral density (PSD) of the white noise\cite{dang2020coordinating}. We can maximize the SNR by maximizing $\|\mathbf{w}\|_1^2/ \|\mathbf{w}\|_2^2$ under some constraints. For the constraints, the Doppler resilience constraint should be still satisfied, i.e.,
\begin{eqnarray}
\mathbf{E}\mathbf{z}=\mathbf{0},	
\end{eqnarray}
where $\mathbf{z} = \mathbf{p}\circ \mathbf{w}$, $p_n \in \{1,-1\}$. Then the optimization is proposed as follows:
\begin{equation}
\begin{array}{ll}
	\displaystyle \max_{\mathbf{w},\mathbf{p}}\, & \frac{\|\mathbf{w}\|_1^2}{\|\mathbf{w}\|_2^2}\\
	 s.t. & \bf{Ez} = \mathbf{0}\\
	\quad & \mathbf{z} = \mathbf{p}\circ \mathbf{w}\\
	\quad & p_n \in \{1,-1\} .
	\end{array}\label{opt:SNR}
\end{equation}
This optimization problem (\ref{opt:SNR}) is a challenging optimization problem because it contains binary variables $p_n$ and complex-valued variables $z_n$. Besides, the objective function in (\ref{opt:SNR}) is not a concave function and the constraint set is not a convex set either.

Because of the difficulty of the proposed optimization problem, we have to transform the complex style into a much simpler form. We will analyze the constraint set first. Suppose $\mathbf{z}_1, \mathbf{z}_2, \cdots,\mathbf{z}_U \in \mathrm{Null}(\mathbf{E})$, and $\bm{\lambda} = [\lambda_1,\lambda_2,\cdots,\lambda_{U}]^T$ is an arbitrary  vector with real number elements, then 
\begin{eqnarray}
\mathbf{Z}\bm \lambda \in \mathrm{Null}(\mathbf{E}).
\end{eqnarray}
i.e.,
\begin{eqnarray}
	\mathbf{E}(\mathbf{Z}\bm \lambda) = \mathbf{0}.
\end{eqnarray}
where $\mathbf{Z} = [\mathbf{z}_1\,\,  \mathbf{z}_2 \,\, \cdots \,\, \mathbf{z}_U]$.
Moreover, for the second and third constraint, it is easy to verify that 
\begin{eqnarray}
	\|\mathbf{Z}\bm \lambda\| = \|\mathbf{p}\circ\mathbf{w}\|=\|\mathbf{w}\|,
\end{eqnarray}
where $\|\cdot\|$ means either $\|\cdot\|_1$ or $\|\cdot\|_2$.

Therefore, the procedure for solving the optimization problem (\ref{opt:SNR}) can be transformed into two steps:
\begin{itemize}
	\item {\bf{Step 1}}. Solve the following 
	\begin{eqnarray}
	\max_{\bm \lambda}\,  \frac{\|\mathbf{Z}\bm{\lambda}\|_1^2}{\|\mathbf{Z}\bm{\lambda}\|_2^2}.\label{opt:SNR1}
\end{eqnarray}

     \item {\bf{Step 2}}. Implement (\ref{eq:p}) and (\ref{eq:w}).
\end{itemize}
\subsection{First Algorithm for SNR}
The optimization problem (\ref{opt:SNR1}) is still a difficult problem, since its objective function is not concave. In order to solve it, two algorithms are proposed. The first algorithm is a simpler one, where the constraint set related to the first algorithm  is limited to a much smaller set than that to the second algorithm. The second algorithm is slightly more difficult than the first algorithm, but the second one can find a better solution. So the first algorithm will be introduced first. Before introducing the first algorithm, a theorem will be proposed.

\emph{Theorem 3:} Let $\lambda_u$ be nonnegative real number with $\sum_{u=1}^{U}\lambda_u=1$, then
\begin{eqnarray}
\max_{\lambda_u} {\|\sum_{u=1}^{U}\lambda_u \mathbf{z}_u\|_1}=  \max_u\{\|\mathbf{z}_u\|_1\}, u=1,2,\cdots,U.
\end{eqnarray}

\begin{proof}	
According to the triangle inequality of norm, i.e.,
\begin{eqnarray}
 \|\sum_{u=1}^{U}\lambda_u \mathbf{z}_u\|_1\leq  \sum_{u=1}^{U}\lambda_u \|\mathbf{z}_u\|_1,
\end{eqnarray}
then 
\begin{eqnarray}
\begin{split}
\max_{\lambda_u} \|\sum_{u=1}^{U}\lambda_u \mathbf{z}_u\|_1 &\leq \max_{\lambda_u} \sum_{u=1}^{U}\lambda_u \|\mathbf{z}_u\|_1\\
&=\max_{u}\{\|\mathbf{z}_u\|_1\}.\label{ineq:right}
\end{split}
\end{eqnarray}

Since $\|\sum_{u=1}^{U}\lambda_u \mathbf{z}_u\|_1 = \|\mathbf{z}_u\|$, when $\lambda_u=1$, $u = 1,2,\cdots,U$, then 
\begin{eqnarray}
	\|\mathbf{z}_v\|_1 \leq \max_{\lambda_u}{\|\sum_{u=1}^{U}\lambda_u \mathbf{z}_u\|_1}, 
\end{eqnarray}
\begin{eqnarray}
	\implies \max_u\{\|\mathbf{z}_u\|_1\} \leq \max_{\lambda_u}{\|\sum_{u=1}^{U}\lambda_u \mathbf{z}_u\|_1}.\label{ineq:left}
\end{eqnarray}

Based on (\ref{ineq:right}) and (\ref{ineq:left}), we can get 
\begin{eqnarray}
\max_{\lambda_u} {\|\sum_{u=1}^{U}\lambda_u \mathbf{z}_u\|_1}=  \max_u\{\|\mathbf{z}_u\|_1\}, u=1,2,\cdots,U.
\end{eqnarray}
\end{proof}

As regards the first method, $\lambda_u$ should be limited by $\lambda_u\geq 0$ and $\sum_{u=1}^{U}\lambda_u=1$. Based on this fact, Theorem 3 implies that $\bm\lambda$ has only one nonzero element, i.e.,1, and the other elements are zeros, which means that the elements of $\bm \lambda$ should be limited to $\sum_{u=1}^{U}\lambda_u=1$ and $\lambda_u\in\{0, 1\}$. Without loss of generality, assume $\mathbf{z}_1,\mathbf{z}_2,\cdots, \mathbf{z}_U$ are normalized, i.e., $\|\mathbf{z}_1\|_2, \|\mathbf{z}_2\|_2, \cdots,\|\mathbf{z}_U\|_2$ are equal to 1, then  we get
\begin{eqnarray}
	\|\mathbf{Z}\bm \lambda\|_2 = 1.
\end{eqnarray}
Therefore, the optimization problem (\ref{opt:SNR1}) is equivalent to 
\begin{eqnarray}
	\max_{\bm \lambda}\,  \|\mathbf{Z}\bm{\lambda}\|_1^2\label{opt:SNR2}
\end{eqnarray}
Without loss of generality, we assume $\|\mathbf{z}_1\|_1\geq \|\mathbf{z}_2\|_1\geq \cdots,\|\mathbf{z}_U\|_1$. Again, based on Theorem 3, the optimal solution to (\ref{opt:SNR2}) is

\begin{equation}
	\lambda_u=\left\{\begin{array}{ll}
1, &\mathrm{if}\,\, u=1\\
0, &\mathrm{otherwise}.\\
\end{array}
\right.
\end{equation}

Let $\hat{\mathbf{z}}=\mathbf{z}_1$, then $\mathbf{p}$ and $\mathbf{w}$ can be obtained from (\ref{eq:p}) and (\ref{eq:w}).

\begin{algorithm}[htb]
\setstretch{1.5}
\caption{Basis selection (BS) method in null space:}
\label{alg:Framwork}
\begin{algorithmic}[1] %
\STATE $\mathbf{z}_1, \mathbf{z}_2, \cdots,\mathbf{z}_U \in \mathrm{Null}(\mathbf{E})$ are the basis vectors.
\STATE Compute $\|\mathbf{z}_1\|_1, \|\mathbf{z}_2\|_1, \cdots,\|\mathbf{z}_U\|_1$.
\STATE if $\|\mathbf{z}_u\|_1$ is the largest one, choose $\mathbf{z}_u$.
\end{algorithmic}
\end{algorithm}

\subsection{Second Algorithm for SNR}
Although the first proposed algorithm has a low computational complexity, it has a very limited performance because the elements of $\bm \lambda$ are restricted to the set $\{0,1\}$. In order to improve the performance, here we proposed a second method --  termed a Heuristic Coordinated Descent (HCD) method where  the binary constraint on $\bm \lambda$ is removed and $\bm \lambda\in \mathbb{C}^U$. HCD can be viewed as a relatively effective method to deal with the non-convex optimization problem with a much larger constraint set. 

Generally speaking, the Coordinate Descent (CD) algorithms\cite{wright2015coordinate} are iterative methods. The most common CD algorithm is by fixing other elements of the variable vector and obtaining the new iteration point by minimizing (maximizing) the objective function with respect to a single element of variable vector. In other words, when an optimization problem was considered, i.e.,
\begin{equation}
	\min_{\bm x\in \mathbb{C}^N} f(\bm x),
\end{equation}
then the CD algorithm starts with some initial vector $\bm{x}^{(0)}=(x_0^{(0)},x_1^{(0)},\cdots,x_{N-1}^{(0)})$ and repeats the following iteration
\begin{eqnarray}
	\begin{aligned}
x_{0}^{(k)} & = \underset{x_{0}}{\operatorname{argmin}} f\left(x_{0},\, x_{1}^{(k-1)},\, x_{2}^{(k-1)},\, \cdots,\, x_{N-1}^{(k-1)}\right), \\
x_{1}^{(k)} & = \underset{x_{1}}{\operatorname{argmin}} f\left(x_{0}^{(k)},\, x_{1},\, x_{2}^{(k-1)},\, \cdots,\, x_{N-1}^{(k-1)}\right), \\
x_{2}^{(k)} & = \underset{x_{2}}{\operatorname{argmin}} f\left(x_{0}^{(k)},\, x_{1}^{(k)},\, x_{2}, \cdots,\, x_{N-1}^{(k-1)}\right), \\
& \vdots \\
x_{N-1}^{(k)} & =\underset{x_{N-1}}{\operatorname{argmin}} f\left(x_{0}^{(k)}, x_{1}^{(k)}, x_{2}^{(k)}, \cdots,\, x_{N-1}\right),
\end{aligned}\label{proc:cd1}
\end{eqnarray}
where $k = 1,2,3,\cdots$.

Since the objective value of (\ref{opt:SNR1}) is always no less than 0, maximizing it is equivalent to minimizing the reciprocal, i.e.,
\begin{eqnarray}
	\min_{\bm \lambda}\,  \frac{\|\mathbf{Z}\bm{\lambda}\|_2^2}{\|\mathbf{Z}\bm{\lambda}\|_1^2}.\label{opt:SNR3}
\end{eqnarray}

 In the optimization problem (\ref{opt:SNR3}), the objective function is still a non-convex function, so that it is difficult for us to obtain the global optimal solution. Based on these difficulties, a heuristic Coordinated Descent (HCD) is proposed. 
 
 The algorithm is based on the CD algorithm. It also starts with some initial vector $\bm{\lambda}^{(0)}=(\lambda_1^{(0)},\lambda_2^{(0)},\cdots,\lambda_n^{(0)})$ and then repeats the procedure as (\ref{proc:cd1}). However, the difference between the HCD and the general CD is that the HCD will revert to the $(k-1)$-th state if the objective value of $k$-th state is higher than  the one from the previous iteration. For details, the following formula shows the  $k$-th itertation of $u$-th element: 
\begin{eqnarray}
\begin{aligned}
\lambda_{u}^{(k)} & = \underset{\lambda_{u}}{\operatorname{argmin}}\,\, f\left(\lambda_{1}^{(k)}, \lambda_{2}^{(k)}, \cdots,\lambda_{u-1}^{(k)},\lambda_{u},\lambda_{u+1}^{(k)}, \cdots, \lambda_{U}^{(k)}\right), \\
\end{aligned}\label{proc:cd1}
\end{eqnarray}
 and then $\lambda_u^{(k)}$ will revert to $\lambda_u^{(k-1)}$ if 
 \begin{eqnarray}
 \begin{split}
  &f\left(\lambda_{1}^{(k)}, \lambda_{2}^{(k)}, \cdots,\lambda_{u-1}^{(k)},\lambda_{u}^{(k)},\lambda_{u+1}^{(k)}, \cdots, \lambda_{U}^{(k)}\right)\\
  &\geq f\left(\lambda_{1}^{(k)}, \lambda_{2}^{(k)}, \cdots,\lambda_{u-1}^{(k-1)},\lambda_{u}^{(k)},\lambda_{u+1}^{(k)}, \cdots, \lambda_{U}^{(k)}\right).\\\label{ineq:g}
  \end{split}
 \end{eqnarray}
 The second difference is that HCD generates many initial vectors. For every initial vector, we repeat the iteration procedure and obtain a solution when it satisfies the stop criteria. For all these initial vectors, we have many solutions from which we can choose the best solution that has the smallest objective value. 
   The algorithm is summarized in algorithm 3. 
 
\begin{algorithm}[htb]
\setstretch{1.5}
\caption{Heuristic Coordinated Descent (HCD) in null space:}
\label{alg:Framwork}
\begin{algorithmic}[] %
\STATE for $i = 1: I$\\
\STATE \quad randomized initial vectors $\bm \lambda^{(0)}$ \\
\STATE \quad for $k = 1:K$\\
\STATE \quad \quad for $u = 1:U$\\
\quad\quad \quad Compute $\lambda_{u}^{(k)}$  as (\ref{proc:cd1})\\
\quad\quad \quad if (\ref{ineq:g}) satisfies, then $\lambda_{u}^{k-1} = \lambda_{u}^{k-1}$.\\
\quad\quad \quad if $\|\bm \lambda^{(k)}-\bm \lambda^{(k-1)} \|_2\leq \varepsilon$; $\hat{\bm \lambda}_i = \bm \lambda^{(k)}$; end\\
\quad\quad end $u$\\
\quad end $k$\\
\STATE end $i$\\
\STATE choose $\hat{\bm \lambda_i}$ as the best $\bm \lambda$ such that (\ref{opt:SNR3}) minimized.
\end{algorithmic}
\end{algorithm}

\section{Windowing for Fully Polarimetric Radar Systems}

The fully polarimetric radar systems can transmit and receive on two orthogonal polarizations at the same time. The use of two orthogonal polarizations increases the degrees of freedom and can result in significant improvement in detection performance.

Two pulse trains $Z_{VP}(t)$ and $Z_{HP}(t)$, transmitted simultaneously from vertical polarization and horizontal polarization, are written by
\begin{eqnarray}
	Z_{VP}(t)=\frac{1}{2} \sum_{n=0}^{N-1}\left[\left(1+p_{n}\right) s_{x}(t-n T)-\left(1-p_{n}\right) \tilde{s}_{y}(t-n T)\right],
\end{eqnarray}
and
\begin{eqnarray}
	Z_{HP}(t)=\frac{1}{2} \sum_{n=0}^{N-1}\left[\left(1-p_{n}\right) \tilde{s}_{x}(t-n T)+\left(1+p_{n}\right) s_{y}(t-n T)\right],
\end{eqnarray}
where $\tilde{\cdot}$ means reversal, i.e.,
\begin{eqnarray}
	\tilde{s}_x(t)=\sum_{l=0}^{L-1}x[L-1-l]u(t-lT_c),\\ \tilde{s}_y(t)=\sum_{l=0}^{L-1}y[L-1-l]u(t-lT_c).
\end{eqnarray}
In the proposed transmission mode, two orthogonal polarizations have different waveforms in a PRI. For example, if $s_x(t)$ (or $-\tilde{s}_{y}(t)$) is transmitted from vertical polarization, then $s_y(t)$ (or $\tilde{s}_{x}(t)$) is transmitted from horizontal polarization. Besides, this mode also contains the famous Alamouti time-space coding when two different waveforms  are transmitted in the adjoint two PRIs . For example, in the $n$-th PRI, horizontal polarization and horizontal polarization transmit $s_x(t)$ and $s_y(t)$ respectively; in the $(n+1)$-th PRI, vertical polarization and horizontal polarization transmit $s_y(t)$ and $\tilde{s}_x(t)$ respectively, which constitudes the famous Alamouti matrix
\begin{equation}
	\left[\begin{matrix}
s_x(t)	& -\tilde{s}_y(t)\\
s_y(t)  & \tilde{s}_x(t)
\end{matrix}\right],
\end{equation}
which can eliminate polarization interference when the target is static. For a Doppler shift resulting in polarization interference, a moving target is considered with a Doppler shift $\omega$ in Hz.

For the vertical polarization antenna, the returned signal is given by
\begin{eqnarray}
	R_V(t) = (h_{VV}Z_{VP}(t)+h_{VH}Z_{HP}(t))e^{j \omega t}.\label{eq:RV}
\end{eqnarray}
Also, similarly, for the horizental polarization antenna, the returned signal is given by
\begin{eqnarray}
	R_H(t) = (h_{HV}Z_{VP}(t)+h_{HH}Z_{HP}(t))e^{j \omega t},\label{eq:RH}
\end{eqnarray}
where $h_{VH}$ denotes the scattering coefficient into the vertical polarization channel from a horizontally polarized incident field\cite{pezeshki2008doppler}. Note that $h_{VV}$, $h_{VH}$, $h_{HV}$, $h_{HH}$ constitute a scattering matrix \begin{equation}
\mathbf{H}=\left[\begin{matrix}
h_{VV} & h_{VH}\\
h_{HV} & h_{HH}
\end{matrix}	
\right].\label{eq:output1}
\end{equation}

At the receiver with two polarization antennas, each antenna has two responses of matched filters, i.e., $Z_{VW}^*(-t)$ and $Z_{HW}^*(-t)$, where
\begin{eqnarray}
\begin{split}
	Z_{VW}(t)=\frac{1}{2} \sum_{n=0}^{N-1}w_n &\left[\left(1+p_{n}\right) s_{x}(t-n T)\right.\\
	 &\left.-\left(1-p_{n}\right) \tilde{s}_{y}(t-n T)\right],\label{eq:ZVW}
\end{split}
\end{eqnarray}
and
\begin{eqnarray}
\begin{split}
	Z_{HW}(t)=\frac{1}{2} \sum_{n=0}^{N-1}w_n &\left[\left(1-p_{n}\right) \tilde{s}_{x}(t-n T)\right.\\
	&\left.+\left(1+p_{n}\right) s_{y}(t-n T)\right].\label{eq:ZHW}
	\end{split}
\end{eqnarray}

The returned signals go through the matched filters then the outputs of the matched filters are given by
\begin{equation}
\mathbf{U}(\tau)=\left[\begin{matrix}
\mathbf{RZ}_{11}(\tau) & \mathbf{RZ}_{12}(\tau)\\
\mathbf{RZ}_{21}(\tau) & \mathbf{RZ}_{22}(\tau)
\end{matrix}	
\right]\label{eq:output1}
\end{equation}
where
\begin{eqnarray}
	\mathbf{RZ}_{11}(\tau)=\int_{-\infty}^{+\infty} R_V(t) Z_{HW}^{*}(t-\tau) d t,\label{eq:RZ11}
\end{eqnarray}

\begin{eqnarray}
	\mathbf{RZ}_{12}(\tau)=\int_{-\infty}^{+\infty} R_V(t) Z_{VW}^{*}(t-\tau)  d t,\label{eq:RZ12}
\end{eqnarray}

\begin{eqnarray}
	\mathbf{RZ}_{21}(\tau)=\int_{-\infty}^{+\infty} R_H(t) Z_{HW}^{*}(t-\tau)  d t,\label{eq:RZ21}
\end{eqnarray}

\begin{eqnarray}
	\mathbf{RZ}_{22}(\tau)=\int_{-\infty}^{+\infty} R_H(t) Z_{VW}^{*}(t-\tau) d t.\label{eq:RZ22}
\end{eqnarray}

After substituting $R_V(t)$ of (\ref{eq:RV}) into (\ref{eq:RZ11}) and  (\ref{eq:RZ12}), and bringing $R_H(t)$ of (\ref{eq:RH}) into (\ref{eq:RZ21}) and (\ref{eq:RZ22}), then we get the following proposition.

\emph{Proposition 3:}
	The output matrix $\mathbf{U}(\tau)$ in (\ref{eq:output1}) can be  transformed into
	\begin{equation}
\mathbf{U}(\tau)=
\left[\begin{matrix}
h_{VV} & h_{VH}\\
h_{HV} & h_{HH}
\end{matrix}	
\right]
\left[\begin{matrix}
\chi_{VP,VW}(\tau,\omega) & \chi_{VP,HW}(\tau,\omega)\\
\chi_{HP,VW}(\tau,\omega) & \chi_{HP,HW}(\tau,\omega)
\end{matrix}	
\right].\label{eq:output2}
\end{equation}	

In (\ref{eq:output2}), the cross ambiguity function $\chi_{a,b}(\tau,\omega)$ is defined as
\begin{eqnarray}
	\chi_{a,b}(\tau, \omega)=\int_{-\infty}^{+\infty} Z_{a}(t) Z_{b}^{*}(t-\tau) \mathrm{e}^{j \omega t} d t,
\end{eqnarray}
where $a = VP \,\, \mathrm{or} \,\, HP$ and $b = VW\,\, \mathrm{or}\,\, HW$.

Based on the transform from (\ref{eq:chiPW}) to (\ref{eq:APW}), the discrete cross ambiguity function $\chi_{VP,VW}(k,\theta)$ is given by
\begin{equation}
	\begin{aligned} \mathcal{A}_{VP, VW}(k, \theta) &=\frac{1}{2}\left[C_{x}(k)+C_{y}(k)\right] \sum_{n=0}^{N-1} w_n e^{j n \theta} \\ &+\frac{1}{2}\left[C_{x}(k)-C_{y}(k)\right] \sum_{n=0}^{N-1} p_{n} w_n e^{j n \theta}. \end{aligned}\label{eq:AVPVW}
\end{equation}

Similarily, the discrete cross ambiguity function $\chi_{VP,VW}(k,\theta)$ is given by
\begin{equation}
	\begin{aligned} \mathcal{A}_{HP, HW}(k, \theta) &=\frac{1}{2}\left[C_{x}(k)+C_{y}(k)\right] \sum_{n=0}^{N-1} w_n e^{j n \theta} \\ &-\frac{1}{2}\left[C_{x}(k)-C_{y}(k)\right] \sum_{n=0}^{N-1} p_{n} w_n e^{j n \theta}.\end{aligned}\label{eq:AHPHW}
\end{equation}

Also, the discrete cross ambiguity function $\chi_{VP,HW}(k,\theta)$ is  given by
\begin{eqnarray}
	&\mathcal{A}_{VP, HW}(k, \theta)=\frac{1}{2}\left[C_{xy}(k)-C_{\tilde{y}\tilde{x}}(k)\right] \displaystyle\sum_{n=0}^{N-1} w_n e^{j n \theta} \nonumber\\
	&\qquad\qquad\qquad\quad+\frac{1}{2}\left[C_{xy}(k)+C_{\tilde{y}\tilde{x}}(k)\right] \displaystyle\sum_{n=0}^{N-1} p_{n} w_n e^{j n \theta}\\
	&\qquad\quad= C_{xy}(k)	\displaystyle\sum_{n=0}^{N-1} p_{n} w_n e^{j n \theta}, \label{AVPHW}			
\end{eqnarray}
and the discrete cross ambiguity function $\chi_{HP,VW}(k,\theta)$ is  given by
\begin{eqnarray}
	&\mathcal{A}_{HP, VW}(k, \theta)=\frac{1}{2}\left[-C_{\tilde{x}\tilde{y}}(k)+C_{yx}(k)\right] \displaystyle\sum_{n=0}^{N-1} w_n e^{j n \theta} \nonumber\\
	&\qquad\qquad\qquad\quad+\frac{1}{2}\left[C_{\tilde{x}\tilde{y}}(k)+C_{yx}(k)\right] \displaystyle\sum_{n=0}^{N-1} p_{n} w_n e^{j n \theta}\\
	&\qquad\quad= C_{yx}(k)	\displaystyle\sum_{n=0}^{N-1} p_{n} w_n e^{j n \theta}. \label{eq:AHPVW}			
\end{eqnarray}

\emph{Proposition 4:}
	After discretization, the output matrix $\mathbf{U}(\tau)$ in (\ref{eq:output2}) can be  transformed into
	\begin{equation}
\mathbf{U}(\tau)=
\left[\begin{matrix}
h_{VV} & h_{VH}\\
h_{HV} & h_{HH}
\end{matrix}	
\right]
\left[\begin{matrix}
\mathcal{A}_{VP,VW}(k,\theta) & \mathcal{A}_{VP,HW}(k,\theta)\\
\mathcal{A}_{HP,VW}(k,\theta) & \mathcal{A}_{HP,HW}(k,\theta)
\end{matrix}	
\right].\label{eq:output3}
\end{equation}

From Proposition 4, to obtain the scattering coefficients,  two conditions must be satisfied:
\begin{itemize}
	\item the range sidelobes of $\mathcal{A}_{VP,VW}(k,\theta)                                                                                                                                                                                                                                                                                                                                                                                                                                                                                                                                                                                                                                                                                                                                                                                                                                                                                                                                 $ and $\mathcal{A}_{HP,HW}(k,\theta)$ should be reduced to zero.
	\item $\mathcal{A}_{VP,HW}(k,\theta)$ and $\mathcal{A}_{HP,VW}(k,\theta)$ should be equal to zero.
\end{itemize}

The range sidelobes of  $\mathcal{A}_{VP,VW}(k,\theta) $ and $\mathcal{A}_{HP,HW}(k,\theta)$ arise from the second terms of (\ref{eq:AVPVW}) and (\ref{eq:AHPHW}) , i.e.,
\begin{eqnarray}
	\frac{1}{2}\left[C_{x}(k)-C_{y}(k)\right] \sum_{n=0}^{N-1} p_{n} w_n e^{j n \theta}.
\end{eqnarray}
Besides,  $\mathcal{A}_{VP,HW}(k,\theta)$ and $\mathcal{A}_{HP,VW}(k,\theta)$ depend on
\begin{eqnarray}
	C_{xy}(k)\sum_{n=0}^{N-1} p_{n} w_n e^{j n \theta} \,\,\mathrm{or}\,\,
	C_{yx}(k)\sum_{n=0}^{N-1} p_{n} w_n e^{j n \theta}.
\end{eqnarray}
In summary, $\mathcal{A}_{VP,VW}(k,\theta) $, $\mathcal{A}_{HP,HW}(k,\theta)$, $\mathcal{A}_{VP,HW}(k,\theta)$ and $\mathcal{A}_{HP,VW}(k,\theta)$ are determined by
\begin{eqnarray}
	\sum_{n=0}^{N-1} p_{n} w_n e^{j n \theta}.
\end{eqnarray}
Therefore, to obtain the scattering coefficients, the following equation must hold:
\begin{eqnarray}
	\sum_{n=0}^{N-1} p_{n} w_n e^{j n \theta}=0, \quad \mathrm{for\,\, all\,}\theta \in \Theta.
\end{eqnarray}
\emph{Theorem 4:}
The range sidelobes of $\mathcal{A}_{VP,VW}(k,\theta)$ are vanished and the value of $\mathcal{A}_{VP, HW}(k, \theta)$ is zero if and only if
	\begin{eqnarray}
		\sum_{n=0}^{N-1} p_{n} w_n e^{j n \theta}=0,  \quad \mathrm{for\,\, all\,}\theta \in \Theta.\label{eq:obj2}
	\end{eqnarray}

Since (\ref{eq:obj2}) is the same as (\ref{eq:obj1}),  then to solve (\ref{eq:obj1}) or (\ref{eq:obj2}) to obtain $\mathbf{p}$ and $\mathbf{w}$, we can use Algorithm 1.

\section{Numerical Results and Discussions}
In this section,  numerical examples are given to verify the results in Sections III and IV. Also, the proposed null space (NS) Doppler resilient scheme and the binomial design (BD) scheme are verified and discussed.

\emph{Remark 4:}
	In some figures, "Amb fcn" is the abbreviation of the ambiguity function.

\subsection{Doppler Resilience in an Interested Doppler Interval for a Single Antenna System}
At first, to show the performance which flexibly suppresses the range sidelobes in the Doppler interval of interest based on the proposed null space algorithm, the specified Doppler interval of interest is given by $\theta \in [0,2]$.   In each continuous Doppler interval, the sampling resolution is $D_I/(M-1)$, where $D_I=2$ and $M=N-1$. Besides, the number of pulses is $N = 48$. Based on the null space algorithm (algorithm 1),  matrix $\mathbf{E}$ is generated as  (\ref{eq:E}), thus the null space of $\mathbf{E}$ can be calculated, and $\mathbf{p}$, $\mathbf{w}$ are also easily obtained based on Algorithm 1. Besides, the Golay  pair is length-64 and is given by
\begin{eqnarray}
\begin{split}
&	\mathbf{x} = [ 1, 1, 1, -1, 1, 1, -1, 1, 1, 1, 1, -1, 1, 1, -1, 1,\\
&\quad\quad 1, 1, 1, -1, -1, -1, 1, -1, 1, 1, 1, -1, -1, -1, 1, -1, \\
&\quad\quad 1, 1, 1, -1, 1, 1, -1, 1, -1, -1, -1, 1, -1, -1, 1, -1, \\
&\quad\quad -1, -1, -1, 1, 1, 1, -1, 1, 1, 1, 1, -1, -1, -1, 1, -1 ],
\end{split}\label{eq:GCPx}
\end{eqnarray}

\begin{eqnarray}
\begin{split}
 &	 \mathbf{y} = [ 1, -1, 1, 1, 1, -1, -1, -1, 1, -1, 1, 1, 1, -1, -1, -1, \\
 &\quad\quad	 1, -1, 1, 1, -1, 1, 1, 1, 1, -1, 1, 1, -1, 1, 1, 1, 1, -1, 1, \\
 &\quad\quad	 1, 1, -1, -1, -1, -1, 1, -1, -1, -1, 1, 1, 1, -1, 1, -1, \\
 &\quad\quad	 -1, 1, -1, -1, -1, 1, -1, 1, 1, -1, 1, 1, 1 ].
\end{split}\label{eq:GCPy}
\end{eqnarray}

\begin{figure}[htbp]
\centering
\subfigure[The value of $p_n$ ]{
\begin{minipage}{7cm}
\centering
\includegraphics[width=1\linewidth]{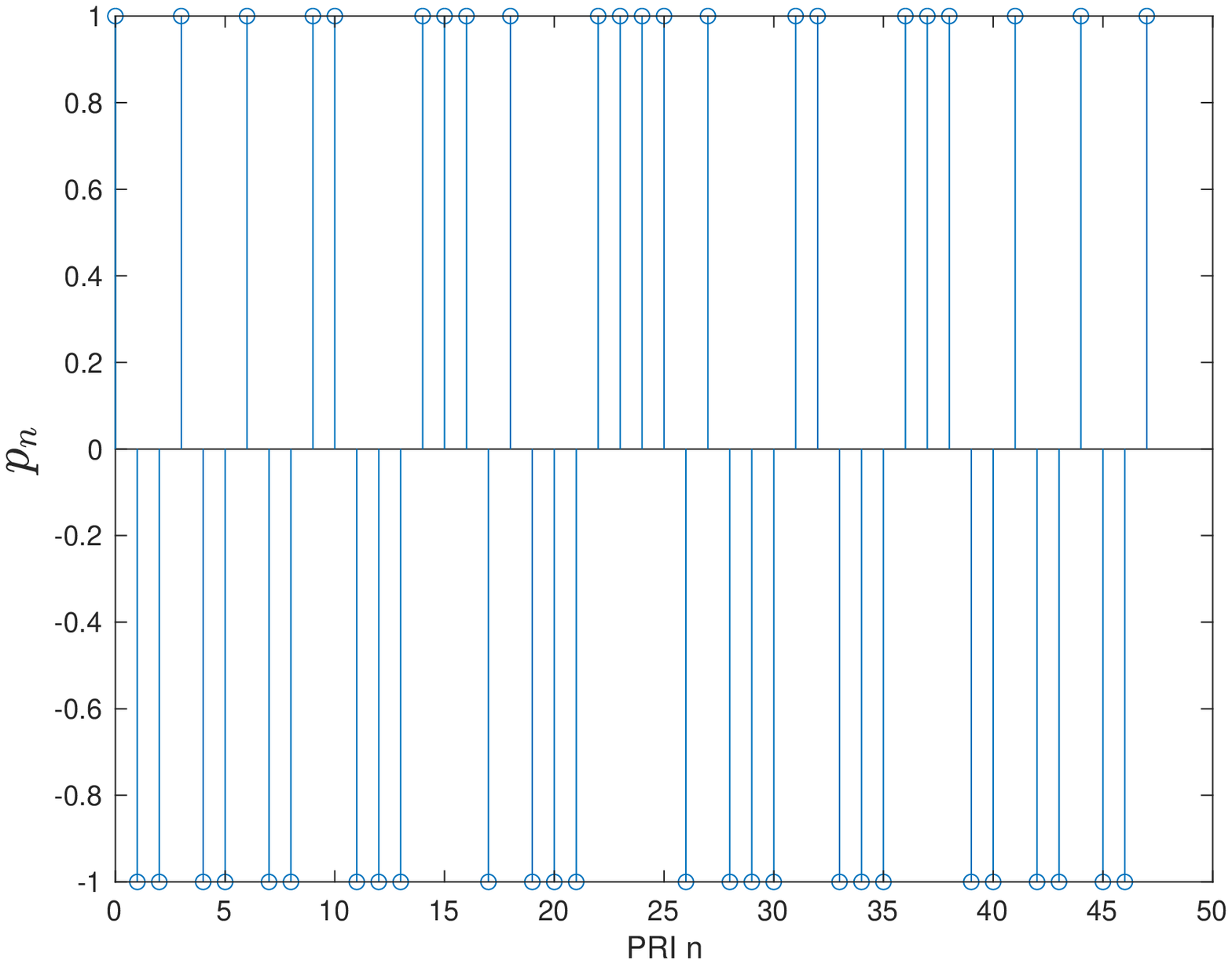}\\
\end{minipage}
}
\subfigure[The modulus of the coefficients of the receiver filter]{
\begin{minipage}{7cm}
\centering
\includegraphics[width=1\linewidth]{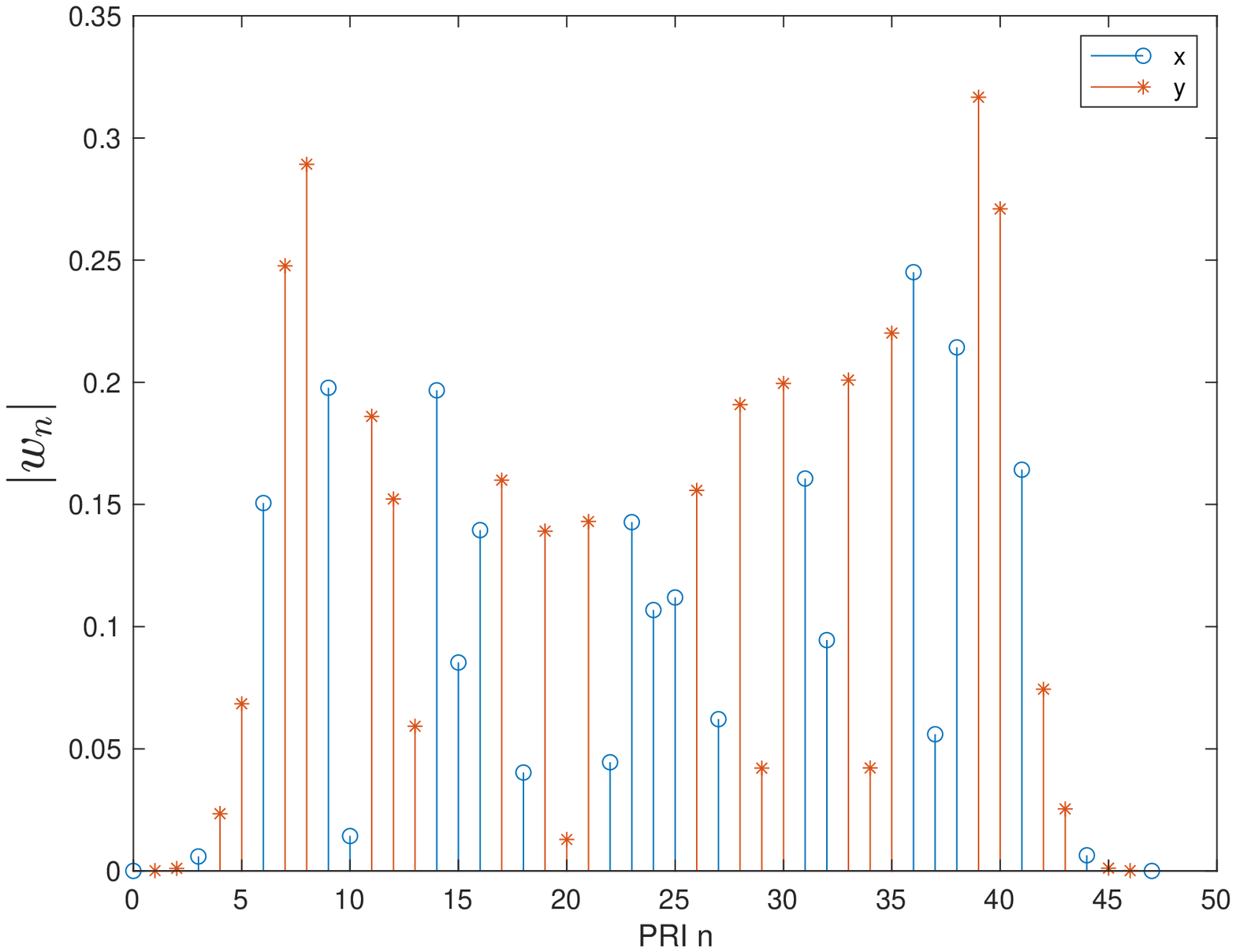}
\end{minipage}
}
\caption{The value of $p_n$ along the PRI $n$ and the modulus  of $w_n$ along the PRI $n$.}
\label{fig1}
\end{figure}

 Fig. \ref{fig1} shows the value of $p_n$ along the PRI $n$ and the modulus of coefficients along the PRI $n$. Then the complementary transmission waveform $Z_P(t)$ in (\ref{eq:Zp}) is determined by $p_n$, and $Z_W(t)$ in (\ref{eq:Zw}) is determined by $p_n$ and $w_n$.

\begin{figure}[htbp]
\centering
\subfigure[AF for the Doppler interval of interest based on Algorithm 1.]{
\begin{minipage}{8cm}
\centering
\includegraphics[width=8cm]{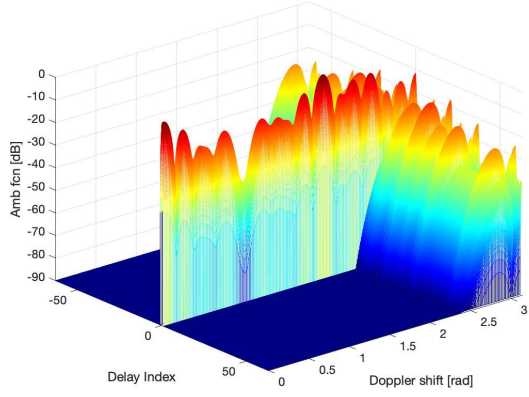}\\
\end{minipage}
}
\subfigure[AF for the Doppler interval of interest based on \cite{chi2009range} and\cite{chi2010complementary}.]{
\begin{minipage}{8cm}
\centering
\includegraphics[width=8cm]{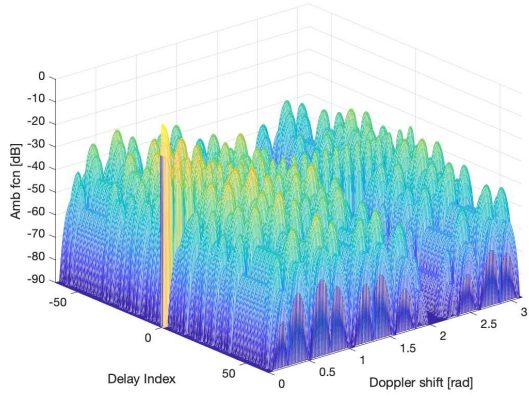}
\end{minipage}
}
\caption{Comparison between (a) AF for the Doppler interval of interest based on algorithm 1, and (b)AF for the Doppler interval of interest based on \cite{chi2009range} and \cite{chi2010complementary}.}
\label{fig2}
\end{figure}

Fig. \ref{fig2} shows two ambiguity functions (\ref{eq:APW}) under  $\theta \in [0,2]$ with $N=48$ based on algorithm 1 and \cite{chi2009range}, respectively. In Fig. \ref{fig2}(a) which is based on Algorithm 1,  the sidelobes within the Doppler interval of interest are obviously lower than the sidelobes outside the Doppler interval of interest. Moreover, in Fig. \ref{fig2}(a), although $[0,2]$ is considerd, the range sidelobes are still very low within $[0,2.4]$.  In Fig. \ref{fig2}(b), based on the oversampled-PTM sequence in \cite{chi2009range}, the ambiguity function has higer range sidelobes than those shown in Fig. \ref{fig2}(a).

\subsection{Doppler Resilience in the Overall Doppler Interval for a Single Antenna System}

We will now discuss the Doppler resilience in the overall Doppler interval $[0,\pi)$. 

\begin{figure}[htbp]
\centering
{
\includegraphics[width=0.8\columnwidth]{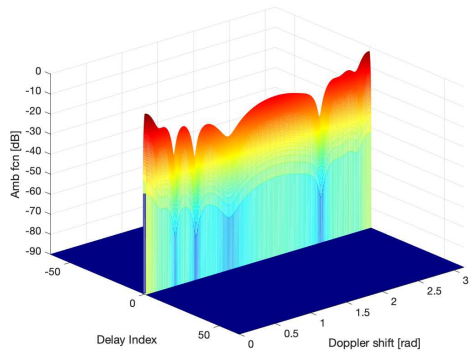}
}
\caption{ AF for overall Doppler interval $[0,\pi)$ based on Algorithm 1.}
\label{fig3}
\end{figure}

\begin{figure}[htbp]
\centering
{
\includegraphics[width=0.8\columnwidth]{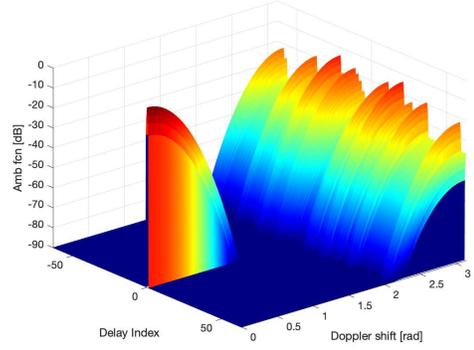}
}
\caption{AF for all Doppler interval $[0,\pi)$ based on the binomial design.}
\label{pic:BigHHN50}
\end{figure}

Fig. \ref{fig3} shows an ambiguity function (\ref{eq:APW}) in the overall Doppler area $ [0,\pi]$ with $N=48$ based on Algorithm 1. For the whole Doppler area, the range sidelobes are lower than -90dB which is an ultra low level.  

Fig. \ref{pic:BigHHN50} is an ambiguity function (\ref{eq:APW}) in the overall Doppler area $ [0,\pi]$ with $N=48$ based on Binomial design, which is a baseline of Fig. \ref{fig3}. In Fig. \ref{pic:BigHHN50}, the range sidelobes gradually increase when the Doppler shift is larger than about 2.2 rad.

From Fig. \ref{fig3} and Fig. \ref{pic:BigHHN50}, it is obvious that the Doppler resilience based on the proposed method is significantly better than that of the BD method in the overall Doppler interval $ [0,\pi]$.

\begin{figure}[htbp]
\centering
{
\includegraphics[width=0.8\columnwidth]{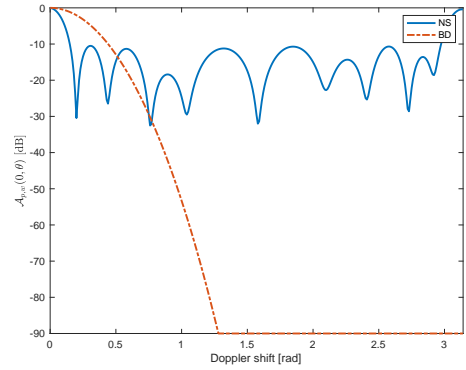}
}
\caption{Doppler profile.}
\label{pic:APW}
\end{figure}

\begin{figure}[htbp]
\centering
{
\includegraphics[width=0.8\columnwidth]{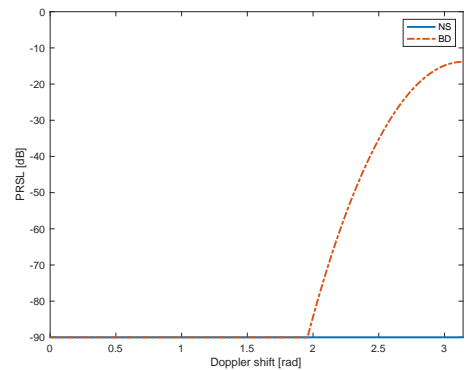}
}
\caption{Peak range sidelobe level. }
\label{pic:PRSL}
\end{figure}

\begin{figure}[htbp]
\centering
{
\includegraphics[width=0.8\columnwidth]{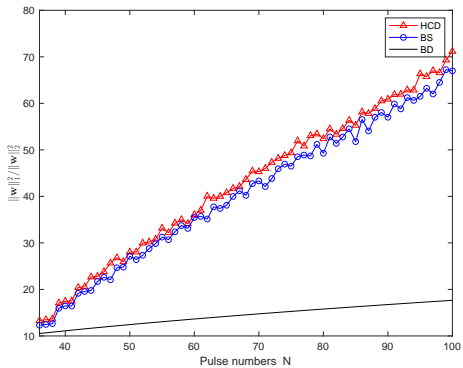}
}
\caption{$\|\mathbf{w}\|_1^2/\|\mathbf{w}\|_2^2$ versus the number of pulses $N$. }
\label{pic:SNR}
\end{figure}

Fig. \ref{pic:APW} and Fig. \ref{pic:PRSL} show the Doppler profile and peak range sidelobe level (PRSL), respectively. From Fig. \ref{pic:APW}, it is observed that the null space (NS) method keeps the mainlobes at a high level but the BD method gradually loses the mainlobe when the Doppler shift increases.  From Fig. \ref{pic:PRSL}, BD also has low sidelobes as well as NS when the Doppler shift is not very big, but when the Doppler increases to a high value, the BD has a high sidelobe level. Besides, from Fig. \ref{pic:PRSL}, the NS has an overall low sidelobe level compared with BD. Moreover, the two figures indicate that the NS method has better Doppler resilience in the overall Doppler interval $[0,\pi]$.

Fig. \ref{pic:SNR} shows that the value of $\|\mathbf{w}\|_1^2/\|\mathbf{w}\|_2^2$ increases when the pulse number $N$ increases. It is obvious that the proposed two methods have a significantly higher SNR than BD. Besides, for the two proposed methods, the heuristic coordinated descent (HCD) slightly outperforms basis selection (BS).

\subsection{Doppler Resilience for Fully Polarimetric Radar Systems}

\begin{figure}[htbp]
\centering
{
\includegraphics[width=0.8\columnwidth]{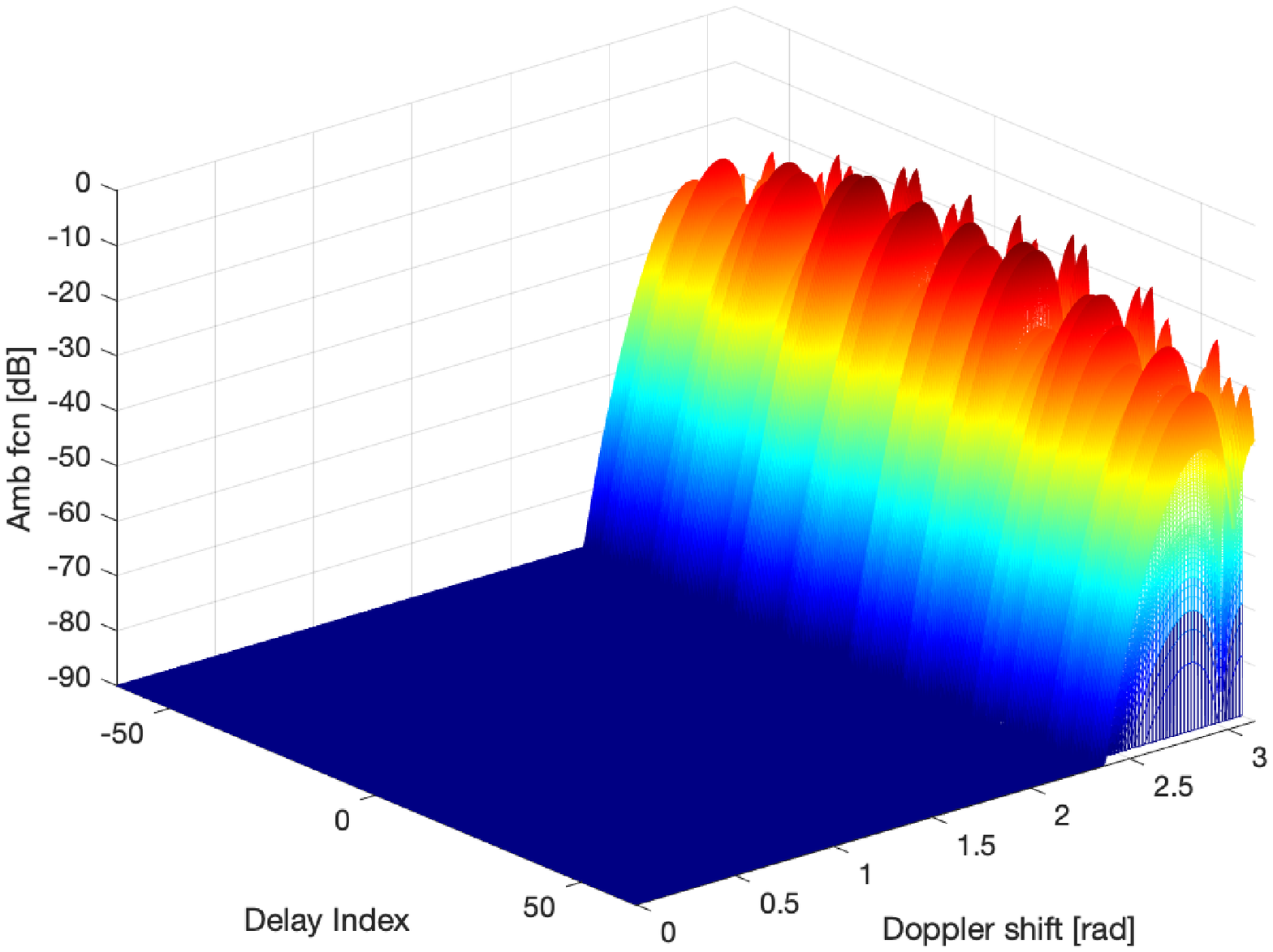}
}
\caption{AF for the intersted Doppler interval about polarimetry interference based on NS.}
\label{pic:AFHVin}
\end{figure}

Based on the analysis in section IV, the transmission scheme is proposed in (\ref{eq:ZVW}) and (\ref{eq:ZHW}). For the vertical polarization antenna (V-antenna), the candidate Golay waveform is $s_x(t)$ and $-\tilde{s}_y(t)$. For the horizontal polarization antenna (H-antenna), the candidate Golay waveforms are $s_y(t)$ and $\tilde{s}_x(t)$. If $s_x(t)$ is transmitted at the V-antenna, then at the same time, $s_y(t)$ should be transmitted. Similarly, if $-\tilde{s}_y(t)$ is transmitted at the V-antenna, then at the same time, $\tilde{s}_x(t)$ should be transmitted. It is noted that here the pulse number $N$ is still 48, and the Golay  pair is shown in (\ref{eq:GCPx}) and (\ref{eq:GCPy}).

To verify the flexible Doppler resilience in fully polarimetric radar systems, the Doppler area is chosen as before, i.e,  $\theta \in [0,2]$. $\mathbf{p}$ and $\mathbf{w}$ are generated via the null space method. After plotting the ambiguity functions $\mathcal{A}_{VP, VW}(k, \theta)$ and $\mathcal{A}_{HP, HW}(k, \theta)$, it is shown that they have the same shape as Fig. \ref{fig2}. Besides, the ambiguity functions $\mathcal{A}_{VP, HW}(k, \theta)$ and $\mathcal{A}_{HP, VW}(k, \theta)$ also have the same shape (shown in Fig. \ref{pic:AFHVin}). From Fig. \ref{pic:AFHVin}, it is observed the values of $\mathcal{A}_{VP, HW}(k, \theta)$ or $\mathcal{A}_{HP, VW}(k, \theta)$ in dB are of a very low level (no more than -90dB) in the interested Doppler area, i.e., $\theta\in [0,2]$. In summary, the range sidelobes and polarimetry interferences are flexibly controlled in the interested Doppler area.

\begin{figure}[htbp]
\centering
{
\includegraphics[width=0.8\columnwidth]{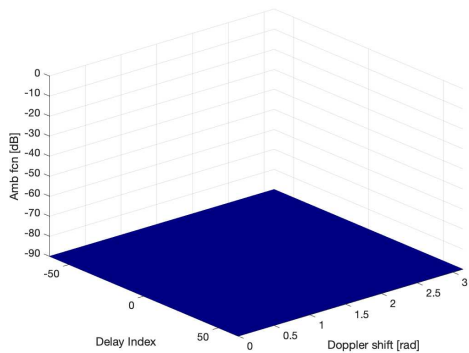}
}
\caption{AF for the overall Doppler interval about polarimetry interference based on NS.}
\label{pic:gHVN50}
\end{figure}

\begin{figure}[htbp]
\centering
{
\includegraphics[width=0.8\columnwidth]{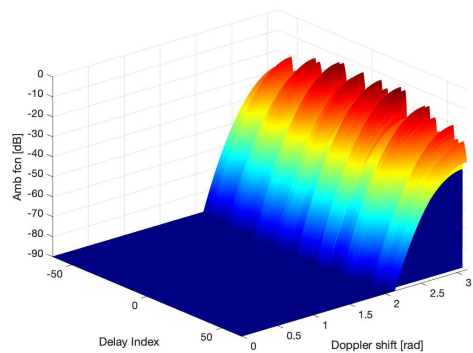}
}
\caption{ AF for the overall Doppler interval about polarimetry interference based on BD.}
\label{pic:BigHVN50}
\end{figure}

Now, the Doppler resilience in the overall Doppler area $[0,\pi]$ is also investigated. After plotting the ambiguity functions $\mathcal{A}_{VP, VW}(k, \theta)$ and $\mathcal{A}_{HP, HW}(k, \theta)$, it is shown that they have the same shape as Fig. \ref{fig3}. Besides, the ambiguity functions $\mathcal{A}_{VP, HW}(k, \theta)$ and $\mathcal{A}_{HP, VW}(k, \theta)$ also have the same shape (shown in Fig. \ref{pic:gHVN50}). From Fig. \ref{pic:gHVN50}, it is observed the values of $\mathcal{A}_{VP, HW}(k, \theta)$ or $\mathcal{A}_{HP, VW}(k, \theta)$ in dB are no more than -90dB, which means that polarimetry interferences are clearly vanished in overall Doppler area $[0,\pi]$.

Thirdly, Fig. \ref{pic:BigHVN50} is a cross ambiguity function generated by BD. It also has excellent low function values, but the values increase when the Doppler shift is bigger than 1 rad, so that it has worse polarimetry interferences as the Doppler increases.   

 \subsection{Discussion}
 Based on the above numerical examples, in which the Doppler resilience is good no matter what the Doppler interval of interest or the overall Doppler interval $[0,\pi)$, one may ask why do we consider the Doppler interval of interest. In fact, a tradeoff exists between the  number of pulses  $N$ and the Doppler interval. 
 
 If the  number of pulses $N$ is too small or the Doppler interval is too big, then $\mathbf{E}$ is a tall matrix which may have full column rank for which the linear system is inconsistent\footnote{It is easy to verify that $\mathbf{E}$ is full rank if $M>N$ and $\theta_{m_1}\not\equiv\theta_{m_2}(\mathrm{mod}2\pi)$, where $m_1,m_2\in \{0,1,\cdots,M-1\}$ and $m_1\neq m_2$. }, so that Algorithm 1 cannot find any solution of the linear system, i.e., $\mathbf{Null}(\mathbf{E})=\varnothing$. Therefore, It makes sense to consider the Doppler interval of interest. 
 
 Moreover. if the  Doppler interval of interest $[0,D_I]$ is given, then the range sidelobes can be suppressed better if the number $M$ of the discrete Doppler shifts $\theta_m \in [0,D_I]$ increases. However, the number $M$ of the discrete Doppler shifts is limited by the number of pulses $N$. This constraint is to ensure the existence of a solution for the addressed linear equations. According to the theory of solutons of linear equations, $\mathbf{Ez}=\mathbf{0}$ has infinite nontrivial solutions if $M<N$. Therefore, $M$ can be chosen as $M = N-1$.

\section{Conclusions}
We have proposed a method based on a null space approach to obtain the Doppler resilient transmission scheme with basic Golay waveforms, which can ensure that the discrete ambiguity function is free of range sidelobes in the Doppler interval of interest or even in overall Doppler interval $[0,\pi)$. Besides, the null space method can be also applied to OFDM signals and obtain the delay resilient OFDM waveform.  Moreover, max-SNR is also considered and optimized by a basis selection method and heuristic coordinate descent methods which are based on the null space. Finally, we have extended the proposed methods to fully polarimetric radar so that range sidelobes and inter-antenna interferences vanish in the overall Doppler interval.


\renewcommand{\refname}{References}
\bibliography{myRef}
\bibliographystyle{myIEEEtran}

\end{document}